\documentclass[%
 amsmath,amssymb,
 reprint,%
]{revtex4-1}
\usepackage{graphicx}
\usepackage{subfigure}
\usepackage{dcolumn}
\usepackage{bm}
\usepackage[utf8]{inputenc}
\usepackage[T1]{fontenc}
\usepackage{mathptmx}
\usepackage{etoolbox}
\usepackage{color}

\makeatletter
\def\@email#1#2{%
 \endgroup
 \patchcmd{\titleblock@produce}
  {\frontmatter@RRAPformat}
  {\frontmatter@RRAPformat{\produce@RRAP{*#1\href{mailto:#2}{#2}}}\frontmatter@RRAPformat}
  {}{}
}%
\makeatother
\begin{document}
\preprint{AIP/123-QED}
\title{Resonant Decay of Kinetic Alfv\'{e}n Waves and 
Implication on Spectral Cascading}
\author{Kexun Shen$^1$}
\author{Zhiwen Cheng$^1$}
\author{Zhiyong Qiu$^{1,2}$} 
\email{E-mail: zqiu@zju.edu.cn}
\affiliation{$^1$Institute for Fusion Theory and Simulation, School of Physics,
Zhejiang University, Hangzhou 310027, China\\
$^2$Center for Nonlinear Plasma Science and 
C.R. ENEA Frascati, C.P. 65, 00044 Frascati, Italy}
\date{\today}

\begin{abstract}

A general equation describing the resonant nonlinear mode-coupling among 
kinetic Alfv\'en waves (KAWs) is derived using nonlinear gyrokinetic theory,  
which can be applied to study the potentially strong spectral energy transfer of KAWs.
As a first application, the parametric decay of a pump KAW into two sideband KAWs 
are studied, with particular emphasis on the cascading in perpendicular wavenumber.  
It is found that, for the ``co-propagating'' cases with all three KAWs propagating
in the same direction along the equilibrium magnetic field line, 
it exhibits a dual cascading character in the perpendicular wavenumber space; 
while for the ``counter-propagating'' cases with one sideband propagating 
in the opposite direction with respect to the pump wave, it instead, 
can exhibit both dual and inverse cascading behaviors. 
The implications on SAW instability nonlinear saturation and charged particle 
transport in fusion plasmas is also discussed.  

\end{abstract}

\maketitle

\section{Introduction} \label{sec:intro}

Shear Alfv\'{e}n waves (SAWs) are incompressible electromagnetic oscillations
prevalent in both nature and laboratory plasmas \cite{HAlfvenNature1942}.
SAWs are often found to mode-convert into small-scale kinetic Alfv\'{e}n waves (KAWs) 
\cite{AHasegawaPRL1975} due to phase-mixing stemming from intrinsic plasma 
nonuniformities \cite{HGradPT1969,LChenPoF1974}. KAWs are uniquely characterized 
by a finite parallel electric field component and 
perpendicular propagation across the mean magnetic field line, which contribute to 
early applications on laboratory plasma heating \cite{AHasegawaPRL1974}, 
geomagnetic pulsation \cite{LChenJGR1974a,LChenJGR1991}, 
and solar corona heating \cite{SRCranmerApJ2003}. 
To quantitatively determine the effects of collisionless plasma transport and 
nonlocal wave energy transfer, one is supposed to acquire detailed 
knowledge of the nonlinear evolution of KAWs,
with comprehensive consideration on both nonlinear wave-wave interactions 
and wave-particle interactions in typical weak turbulence theory \cite{RZSagdeev1969}. 
Resonant three-wave  interaction \cite{LStenfloPS1994}, in this respect,
is a basic process resulting in various nonlinear wave dynamic evolution. 
Previous studies on the nonlinear mode coupling process of SAW/KAW have focused on 
the resonant parametric decay into ion acoustic wave and/or ion-induced scattering, 
where qualitative difference between 
ideal-magnetohydrodynamic (MHD) results \cite{RZSagdeev1969} and 
kinetic results \cite{LChenEPL2011,AHasegawaPoF1976} are addressed, which 
demonstrate the crucial importance of kinetic description in 
the study of KAW nonlinear processes. 

Magnetized plasma turbulence in both space and fusion devices 
are very often, constituted by fluctuations with frequency $\omega$  
much lower than the ion cyclotron frequency $\Omega_i$, 
and anisotropy in directions  parallel and perpendicular to the magnetic field. 
The parallel wavelength can be up to the system size, while the perpendicular 
wavelength varying from system size to ion Larmor radius $\rho_i$.
One example is the drift wave (DW) type micro-turbulence excited by expansion free 
energy associated by plasma nonuniformities, and its nonlinear dynamics 
including spectral evolution in the strong turbulence limit can be described by 
the famous Charney-Hasegawa-Mima (CHM) equation \cite{AHasegawaPRL1977,AHasegawaPoF1978}.
Charney-Hasegawa-Mima equation can also describe quasi-two-dimensional turbulence in 
atmospheric motion of a rotating planet \cite{PBRhinesJFM1975}, 
and it reveals essential features of conservation constraints like energy and enstrophy 
with respective cascading behaviors and self-organization processes 
\cite{AHasegawaAP1985} in analogy to two-dimensional neutral viscid fluid system 
described by the Navier-Stokes equation \cite{GBoffettaARFM2012,RHKraichnanRPP1980}.
From the dispersion relation of Alfv\'{e}n waves, with predominantly 
$\omega\propto k_{\parallel}$ to the leading order and $k_{\parallel}$ being the 
wavenumber parallel to the equilibrium magnetic field, one can physically speculate 
that strong decays and coalescences could occur since the wavenumber and frequency 
matching conditions required for resonant three-wave interactions can be easily 
satisfied for Alfv\'{e}n waves.  
Thus, this kind of self-interaction mechanism and strong coupling could result in 
high turbulent level and broad energy spectrum, which is 
universal in various systems and 
commonalities shall exist for Alfv\'{e}nic turbulence with 
practical interest in solar wind 
\cite{TSHorburyPPCF2005,FSahraouiPRL2009,CSSalemApJL2012},
interstellar medium \cite{JWArmstrongApJ1995,AHMinterApJ1996} 
and accretion disks \cite{EQuataertApJ1998,EQuataertApJ1999}. 
Generally speaking, the standard turbulence paradigm 
\cite{UFrischCUP1995,JAKrommesARFM2012,AASchekochihinPPCF2008}
involves long-wavelength energy-containing scales, 
inner dissipation scales, and disordered fluctuations 
filling up a broad range of intermediate scales (i.e., the inertial range), 
it is thus of necessity to adopt the nonlinear gyrokinetic theory
\cite{AJBrizardRMP2007,EAFriemanPoF1982} with the anisotropic 
assumption to fully describe Alfv\'{e}nic turbulent cascading 
\cite{AASchekochihinApJS2009,GGHowesApJ2006}
both analytically and numerically with arbitrary spatial resolution. 
In the nonlinear gyrokinetic equation, 
we see that particle distribution functions in the gyrocenter phase space 
are nonlinearly phase-mixed by the gyroaveraged electric-field drift induced 
convective flows \cite{AASchekochihinPPCF2008},
which brings the energy injected at the outer scale down to collisional dissipations 
at particle gyroscales. The analytics shall further lead to concrete predictions 
for the spectra, conservative laws, statistical properties, ordered structures 
and self-consistent states of Alfv\'{e}nic turbulence.

Additionally, since SAW instabilities could be resonantly excited by 
energetic charged particles (EPs) \cite{AFasoliNF2007,LChenRMP2016} 
as discrete Alfv\'{e}n eigenmodes (AEs) 
or energetic particle continuum modes \cite{LChenRMP2016} in fusion plasmas
characterized by multiple short-wavelength 
$|k_{\perp}\rho_i|\sim\mathcal{O}(10^{-1})$ modes, 
with $k_{\perp}$ being the wavenumber perpendicular to the equilibrium 
magnetic field, the nonlinear interactions among SAW/KAW triplets can then 
transfer energy from  a linearly unstable primary mode to stable modes, providing
a fundamental nonlinear saturation mechanism for SAW instabilities 
in burning plasmas, which has already been observed in 
simulations \cite{PLiuPRL2022,PLiuRMPP2023,LYeNF2023} and analysed theoretically 
\cite{SWeiNF2022}. The analysis on KAW/SAW nonlinear interactions, can thus be 
applied to study the spectral transfer among various AEs in fusion plasmas, 
with the ultimate goal of understanding the confinement of EPs and plasma performance
in future fusion reactors.

In this work,  we mainly address the potential spectral cascading behaviors, 
by deriving the general nonlinear equation describing resonant three-wave interactions
among KAWs, and studying the parametric decay of a pump KAW, with emphasis on
the condition for the process to spontaneously occur. 
The rest of the manuscript is organized as follows. In Sec. \ref{sec:model}, 
the gyrokinetic theoretical framework is introduced, which is used in 
Sec. \ref{sec:NLeqn} to derive the general nonlinear equation describing 
resonant three-KAW interactions. The parametric decay of a pump KAW is analyzed 
in Sec. \ref{sec:PDI}, with emphasis on the condition for spontaneous decay. 
Finally the results  are summarized in Sec. \ref{sec:conclu}, 
where future work along this line is also briefly discussed. 

\section{Model equations} \label{sec:model}

Considering a simple slab geometry and uniform plasma, we investigate the resonant 
nonlinear interaction among three KAWs in low-$\beta$ magnetized plasmas using 
nonlinear gyrokinetic theory \cite{AJBrizardRMP2007,EAFriemanPoF1982}. The magnetic 
compression is systematically suppressed by the $\beta\lesssim \mathcal{O}(10^{-1})$
and $k_{\parallel}\ll k_{\perp}$ orderings of interest. Here, 
$\beta =8\pi P/B^2_0$ is the ratio of thermal to magnetic pressure,  
$P$ is the plasma thermal pressure, $B_0$ is the equilibrium magnetic field amplitude. 
Although the existence of KAW is related to intrinsic plasma nonuniformities 
\cite{AHasegawaPRL1974,AHasegawaPRL1975} and/or wave-particle interactions 
\cite{LChenRMP2016,AFasoliNF2007,YTodoRMPP2018}, we, for the clarity of discussion, 
only focus on the essential physical mechanism of KAW resonant decay and the 
implication on spectrum cascading assuming uniform plasmas, while their linear stability,
important for the spectrum cascading and final saturation, are not accounted for here.
The effects of magnetic field geometry, plasma nonuniformities, and/or trapped 
particle effect, potentially impacting the KAW decay in magnetically confined 
high-temperature plasmas, are also neglected \cite{LChenPoP2013}. Following the 
standard approach \cite{EAFriemanPoF1982}, the perturbed distribution function 
$\delta f_j$ for species $j=i,e$ is given by
\begin{equation} 
  \delta f_j=- (q/T)_jF_{Mj}\delta\phi 
  +e^{-\bm{\rho}_j\cdot\bm{\nabla}}\delta g_j,
\end{equation}
with $q_j$ being the particle's charge, 
$F_{Mj}$ and $T_j$ being the local Maxwellian distribution function
and the equilibrium temperature respectively,
$\delta\phi=\sum_{\bm{k}}[\delta\phi_k(t)e^{i\bm{k}\cdot\bm{r}}+c.c.]/2$,
and $e^{-\bm{\rho}_j\cdot\bm{\nabla}}$ denoting 
the generator of coordinate transformation from 
guiding-center space to particle phase space.
The nonadiabatic particle response $\delta g_j$ can be derived from the 
nonlinear gyrokinetic equation
\begin{eqnarray} 
  \left[\partial_t+v_{\parallel}\hat{\bm{b}}\cdot\bm{\nabla}
  +(c/B_0)\hat{\bm{b}}\times\bm{\nabla}\langle\delta L_{g,j}\rangle_{\alpha}
  \cdot\bm{\nabla}\right]\delta g_j \nonumber \\ 
  =\left(q/T\right)_jF_{Mj}\partial_t\langle\delta L_{g,j}\rangle_{\alpha},
\end{eqnarray}
with $\langle (...)\rangle_{\alpha}$ denoting gyro-averaging
and $\hat{\bm{b}}=\bm{B}_0/B_0$. 
The leading-order nonlinear $E\times B$ convection term is represented 
by the effective gyroaveraged potential
$\langle\delta L_{g,j}\rangle_{\alpha}\equiv
\langle \exp (\bm{\rho}_j\cdot\bm{\nabla})
(\delta\phi -v_{\parallel}\delta A_{\parallel}/c)\rangle_{\alpha}$,
which  includes the contribution of 
both the perturbed electric-field drift term 
$(c/B_0)\hat{\bm{b}}\times\bm{\nabla}_{\perp}\delta\phi$
and the magnetic flutter term 
$(v_{\parallel}/B_0)\bm{\nabla}_{\perp}\delta A_{\parallel}\times\hat{\bm{b}}$.
Assuming thermal ion species with unit electric charge $e$ and particle density $n_0$,
the governing field equations are the quasi-neutrality condition
\begin{equation}\label{eqn:QNcond} 
  (1+\tau)\delta\phi_k =T_e/(n_0e)\langle J_k\delta g_{k,i}-\delta g_{k,e}\rangle_v,
\end{equation}
and the nonlinear gyrokinetic vorticity equation \cite{LChenJGR1991,LChenNF2001}
\begin{eqnarray}\label{eqn:NLGKV} 
  && ik_{\parallel}\delta J_{\parallel k}+(n_0e^2/T_i)(1-\Gamma_k)\partial_t\delta\phi_k
  \nonumber \\ 
  && =\sum_{\bm{k}=\bm{k}'+\bm{k}''}(\Lambda_{k''}^{k'}/2)
  [\delta A_{\parallel k'}\delta J_{\parallel k''}/c \nonumber \\ 
  && -e\langle (J_kJ_{k'}-J_{k''})\delta L_{k'}\delta g_{k'',i}\rangle_v],
\end{eqnarray}
derived from the parallel Ampere's law 
$\delta J_{\parallel k}=(c/4\pi)k_{\perp}^2\delta A_{\parallel k}$,
the quasi-neutrality condition, and the nonlinear gyrokinetic equation. 
Here, $\tau\equiv T_e/T_i$ is the temperature ratio, $\langle (...)\rangle_v$ 
denotes the velocity-space integration
and $J_k(k_{\perp}\rho)=\langle \exp (\bm{\rho}\cdot\bm{\nabla})\rangle_{\alpha}$ 
represents the finite-Larmor-radius (FLR) correction.
Furthermore, $\Gamma_k=\langle J_0^2F_{Mi}/n_0\rangle_v=I_0(b_k)\exp (-b_k)$ 
with $I_0$ being the modified Bessel function, 
$b_k=k_{\perp}^2\rho_i^2$, $\rho_i=v_{ti}/\Omega_i$, $v_{ti}=\sqrt{T_i/m_i}$,
$\Lambda_{k''}^{k'}\equiv (c/B_0)\hat{\bm{b}}
\cdot\bm{k}_{\perp}''\times\bm{k}_{\perp}'$ 
represents nonlinear perpendicular scattering
with the wavenumber matching condition $\bm{k}=\bm{k}'+\bm{k}''$ applied, and 
$J_k(k_{\perp}\rho_e)\simeq 1$ due to $|k_{\perp}\rho_e|\ll 1$.
The   two terms on the left-hand-side of Eq. (\ref{eqn:NLGKV})
represent field line bending and plasma inertia, respectively; 
while the terms on the right-hand-side of Eq. (\ref{eqn:NLGKV}) are the formally 
nonlinear terms from the Maxwell stress (MX) and 
generalized gyrokinetic Reynolds stress (RS) contribution.
It is noteworthy  that present nonlinear terms will be dominated  by the polarization 
nonlinearity in the limit of $|k_{\perp}\rho_i|^2\ll |\omega /\Omega_i|$
\cite{LChenEPL2011,LChenPoP2013,AHasegawaPoF1976},
which has significant role in nonlinear MHD description 
\cite{TSHahmPRL1995,RZSagdeev1969}.
Therefore, the analyses of the present work are valid 
for $|k_{\perp}\rho_i|^2/|\omega /\Omega_i|\gtrsim\mathcal{O}(1)$ parameter regime, 
which should be kept in mind as we proceed to the investigation of cascading behaviors 
in future works.

\section{Nonlinear Mode Equation} \label{sec:NLeqn}

Now we proceed to derive the governing equations describing resonant three-KAW interaction. 
For a KAW $\bm{\Omega}_k$ with frequency $\omega_k$ and 
wavenumber $\bm{k}=k_{\parallel}\hat{\bm{b}}+\bm{k}_{\perp}$, 
strong scattering could occur for each pair of sidebands 
$\bm{\Omega}_{k'}=(\omega_{k'},\bm{k}')$
and $\bm{\Omega}_{k''}=(\omega_{k''},\bm{k}'')$ with 
the frequency/wavenumber matching condition 
$\bm{\Omega}_k=\bm{\Omega}_{k'}+\bm{\Omega}_{k''}$ satisfied.
Without loss of generality, we assume that $v_{te}\gg |\omega_k/k_{\parallel}|\gg v_{ti}$ 
for all Fourier modes involved, and the linear particle responses of KAWs are derived as
\begin{eqnarray}
    \delta g_{k,i}^{(1)}&\simeq& \frac{e}{T_i}F_{Mi}J_k\delta\phi_k,\\
    \delta g_{k,e}^{(1)}&\simeq& -\frac{e}{T_e}F_{Me}\delta\psi_k.
\end{eqnarray}
Here, the effective potential 
$\delta\psi_k =\omega_k\delta A_{\parallel k}/(ck_{\parallel})$ 
corresponding to the induced parallel electric field is introduced, and 
$\delta\psi_k=\delta\phi_k$ is equivalent to the ideal MHD condition 
($\delta E_{\parallel}=0$).
Substituting linear particle responses into quasi-neutrality condition, one obtains
\begin{equation}
    \delta\psi_k =\sigma_k\delta\phi_k,
\end{equation}
which then yields, together with linear gyrokinetic vorticity equation,
the linear dielectric function of KAW
\begin{equation}
    \epsilon_{Ak}=(1-\Gamma_k)/b_k-(k_{\parallel}^2v_A^2/\omega_k^2)\sigma_k, \label{eq:KAW_DR}
\end{equation}
with $\sigma_k=1+\tau (1-\Gamma_k)$ indicating the deviation from ideal MHD constraint 
due to FLR effect and generation of finite parallel electric field that is crucial for 
plasma heating. Furthermore, $v_A=B_0/\sqrt{4\pi n_0m_i}$ is the Alfv\'{e}n speed.
We note that, only the Hermitian part of KAW dielectric function is 
given in Eq. (\ref{eq:KAW_DR}) for clarity of notation, while its anti-Hermitian part 
for linear stability crucial for later analysis of KAW parametric decay and spectrum
cascading can be recovered straightforwardly. Eq. (\ref{eq:KAW_DR}) yields, 
the familiar expression of KAW dispersion relation in the $b_k\ll1$ limit
\begin{eqnarray}
  \omega_k^2\simeq k^2_{\parallel}v_A^2\left(1+k^2_{\perp}\rho^2_{\kappa}\right),
\end{eqnarray}
with $\rho^2_{\kappa}\equiv (3/4+\tau)\rho^2_i$.  The relation between field potentials simultaneously gives the polarization 
properties of KAW as 
\begin{equation}
    c|\delta\bm{E}_{\perp}|/|\delta\bm{B}_{\perp}|=v_A\sqrt{b_k/[\sigma_k(1-\Gamma_k)]},
\end{equation}
and
\begin{equation}
    c|\delta E_{\parallel}|/|\delta\bm{B}_{\perp}|=
    v_A|k_{\parallel}/k_{\perp}|\tau\sqrt{b_k(1-\Gamma_k)/\sigma_k},
\end{equation}
which suggest the difference between energy spectra of 
perpendicular electric and magnetic perturbations
in both inertial and dissipation ranges of Alfv\'{e}nic 
solar wind turbulence \cite{SDBalePRL2005}.
It has been demonstrated that even in a fully developed turbulence, 
the fluctuations could retain 
the bulk of linear physics of KAWs \cite{GGHowesPRL2008,GGHowesPRL2011}.
 
The nonlinear electron response to $\Omega_k$ can be derived from 
the nonlinear gyrokinetic equation as
\begin{eqnarray}
  \delta g_{k,e}^{(2)}\simeq && \sum_{\bm{k}=\bm{k}'+\bm{k}''} 
  (i\Lambda_{k''}^{k'}/2k_{\parallel})(eF_{Me}/T_e) \nonumber \\ 
  && \times (k_{\parallel}'/\omega_{k'}-k_{\parallel}''/\omega_{k''})
  \delta\psi_{k'}\delta\psi_{k''},
\end{eqnarray}
while the nonlinear ion response is negligible due to 
$\sqrt{\beta_i}\sim v_{ti}/v_A\ll 1$. 
Applying the quasi-neutrality condition, i.e. Eq. (\ref{eqn:QNcond}), we obtain 
\begin{eqnarray}\label{eqn:QNcond2}
  \delta\psi_k= && \sigma_k\delta\phi_k+\sum_{\bm{k}=\bm{k}'+\bm{k}''}
  (i\Lambda_{k''}^{k'}/2k_{\parallel}) \nonumber \\ 
  && \times (k_{\parallel}'/\omega_{k'}-k_{\parallel}''/\omega_{k''})
  \delta\psi_{k'}\delta\psi_{k''},
\end{eqnarray}
which indicates that the coupling of $\Omega_{k'}$ and $\Omega_{k''}$ 
due to electron responses 
could nonlinearly contribute to additional parallel electric field perturbation.
The nonlinear gyrokinetic vorticity equation, Eq. (\ref{eqn:NLGKV}), on the other hand,  yields
\begin{eqnarray}\label{eqn:NLGKV2}
  && b_k\left(\frac{1-\Gamma_k}{b_k}\delta\phi_k-
  \frac{k_{\parallel}^2v_A^2}{\omega_k^2}\delta\psi_k\right)\nonumber \\ 
  && =-\sum_{\bm{k}=\bm{k}'+\bm{k}''}(i\Lambda_{k''}^{k'}/2\omega_k)
  [(\Gamma_{k'}-\Gamma_{k''})\delta\phi_{k'}\delta\phi_{k''} \nonumber \\
  && +(k_{\parallel}'k_{\parallel}''/\omega_{k'}\omega_{k''})
  (b_{k'}-b_{k''})v_A^2\delta\psi_{k'}\delta\psi_{k''}].
\end{eqnarray}
Substituting Eq. (\ref{eqn:QNcond2}) into Eq. (\ref{eqn:NLGKV2}), one then obtains
the desired nonlinear equation for KAW resonant three-wave interactions     
\begin{equation}\label{eqn:NLEqn}
  b_k\epsilon_{Ak}\delta\phi_k=-\sum_{\bm{k}=\bm{k}'+\bm{k}''}
  (i\Lambda_{k''}^{k'}/2\omega_k)\beta_{k',k''}\delta\phi_{k'}\delta\phi_{k''},
\end{equation}
with the nonlinear coupling coefficient expressed as
\begin{eqnarray}\label{eqn:NLCoupCoeff}
  \beta_{k',k''} = && (\Gamma_{k'}-\Gamma_{k''})+
  \frac{k_{\parallel}'k_{\parallel}''(b_{k'}-b_{k''})v_A^2}{\omega_{k'}\omega_{k''}}
  \sigma_{k'}\sigma_{k''} \nonumber \\ 
  && -b_kv_A^2\frac{k_{\parallel}}{\omega_k}
  \left(\frac{k_{\parallel}'}{\omega_{k'}}-\frac{k_{\parallel}''}{\omega_{k''}}\right)
  \sigma_{k'}\sigma_{k''}.
\end{eqnarray}

Eq. (\ref{eqn:NLEqn}) describes nonlinear dynamics of low-frequency 
KAW due to resonant three-KAW interactions in uniform systems, 
and is the most important result of the present work. 
The present form of the nonlinear coupling coefficient given by  
Eq. (\ref{eqn:NLCoupCoeff}) allows us to 
analyze the contribution from different terms with clear  physical meanings.
The first two terms represent the contribution of RS and 
MX, both of which are negligible in the long-wavelength limit
\cite{LChenEPL2011},  and the  last term originates from finite parallel electric field contribution 
via the field line bending term in vorticity equation.

We shall further defining $s_k$ as the direction of a given mode $(\omega_k,\bm{k})$  
propagation along the magnetic field line, with the value of either $-1$ or $+1$,  
and $\alpha_k\equiv|\omega_k/(k_{\parallel}v_A)|=\sqrt{\sigma_kb_k/(1-\Gamma_k)}$ 
as the modulus of $\omega_k/(k_{\parallel}v_A)$.
Meanwhile, $\alpha_k$ increases with the perpendicular wavenumber monotonically. 
The nonlinear coupling coefficient $\beta_{k',k''}$ can, thus, be re-written as
\begin{eqnarray}
  \beta_{k',k''}= && \sigma_{k'}\sigma_{k''}
  \left(\frac{s_{k''}}{\alpha_{k''}}-\frac{s_{k'}}{\alpha_{k'}}\right)\nonumber \\
  && \times\left(b_k\frac{s_k}{\alpha_k}+b_{k'}\frac{s_{k'}}{\alpha_{k'}}
  +b_{k''}\frac{s_{k''}}{\alpha_{k''}}\right).
\end{eqnarray}
It is noteworthy that similar expression of 
the nonlinear coupling coefficient $\beta_{k',k''}$
was also derived in Ref. \citenum{YMVoitenkoJPP1998}, 
where the nonlinear mode equation is derived via integral 
along unperturbed particle orbit with the low-frequency ($|\omega /\Omega_i|\ll 1$) limit.
The propagation direction of each mode, thus, crucially
determines the magnitude of nonlinear coupling coefficient.
Expanding $\beta_{k',k''}$ up to the order of $\mathcal{O}(b_k)$ 
in the  $b_k\ll 1$ limit,
we obtain 
\begin{equation}
  \beta_{k',k''}\simeq (1-s_{k'}s_{k''})(b_{k''}-b_{k'})-s_k(s_{k'}-s_{k''})b_{k},
\end{equation}
which indicates negligible coupling  as two sidebands are co-propagating 
($s_{k'}s_{k''}=1$), 
while retain finite coupling for counter-propagating sidebands ($s_{k'}s_{k''}=-1$)
due to additive effect of RS, MX and 
finite nonlinear parallel electric field contribution. Therefore,
higher-order FLR correction $\mathcal{O}(b_k^2)$ should be considered 
for co-propagating case and leads to 
\begin{equation}
  \beta_{k',k''}^{\text{co-}}\simeq (3/8+\tau /2)(b_{k'}-b_{k''})(b_k+b_{k'}+b_{k''}).
\end{equation}
Similarly, we can derive the simplified expression of $\beta_{k''}^{k'}$
in the short-wavelength limit $b_k\gg 1$ as 
\begin{eqnarray}
  \beta_{k',k''}\simeq && (1+\tau)\left(\frac{s_{k''}}{\sqrt{b_{k''}}}
  -\frac{s_{k'}}{\sqrt{b_{k'}}}\right) \nonumber \\ 
  && \times \left(s_k\sqrt{b_k}+s_{k'}\sqrt{b_{k'}}+s_{k''}\sqrt{b_{k''}}\right).
\end{eqnarray}
These analytical expressions shall be used as verification for subsequent resultant cascading  behaviors.

As a brief remark, the nonlinear gyrokinetic equation and vorticity equation 
applied here are able to capture the crucial  features of 
the low-frequency turbulent phenomena and 
depict the nonlinear behaviors of KAW turbulence in great detail.
Similarity between the Charney-Hasegawa-Mima equation
and the given Eq. (\ref{eqn:NLEqn}) is evident 
since both electrostatic low-frequency drift wave turbulence and 
KAW turbulence could be characterized as strong turbulence with 
finite-amplitude fluctuations and broadband spectra in $\omega$ and $\bm{k}_{\perp}$.
However, the more complicated form of nonlinear coupling coefficient indicates potentially 
richer turbulent phenomena for KAW turbulence. 
Furthermore, current analysis can be straightforwardly extended to account for
plasma nonuniformities  to 
describe general drift-Alfv\'{e}n turbulence \cite{LChenNF2001} in laboratory  plasmas.
At the present stage, to picture the universal spectra of KAW turbulence 
in both laboratory and nature plasmas, 
we shall investigate the potential turbulent cascading properties,  
using the paradigm model of a KAW resonantly decaying into two KAW sidebands. 

\begin{figure}[htbp!]
  \centering
  \subfigure{\label{fig:1a}\includegraphics[width=.9\linewidth]{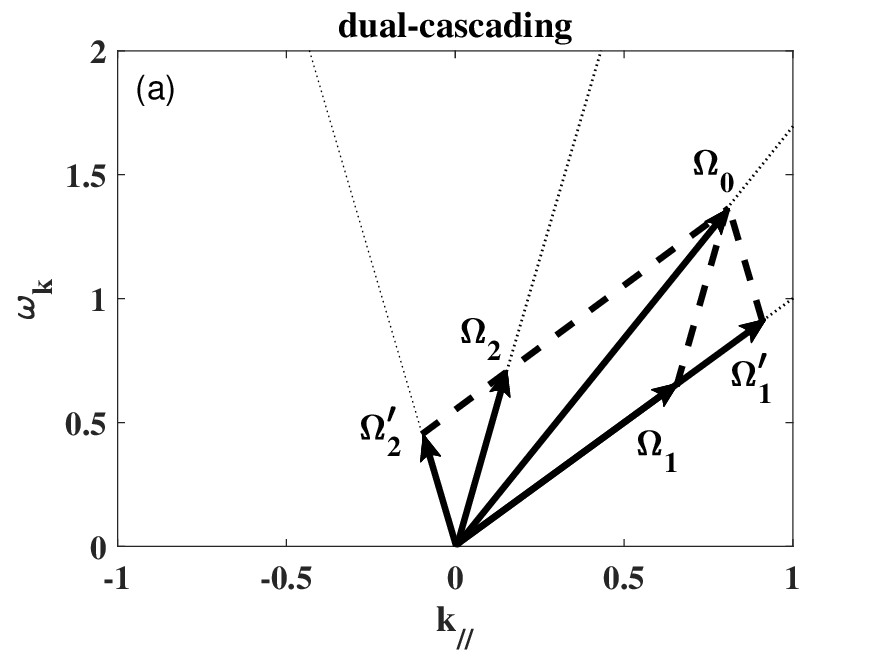}}
  \subfigure{\label{fig:1b}\includegraphics[width=.9\linewidth]{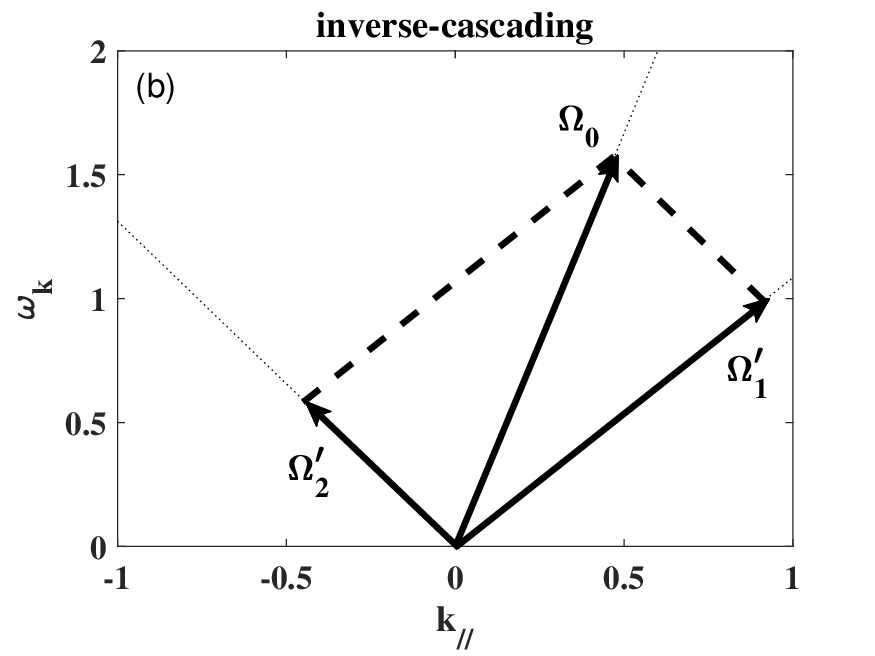}}
  \caption{Sketches of frequency and parallel wavenumber matching conditions
  ($\omega_{k_0}=\omega_{k_1}+\omega_{k_2}$, 
  $k_{\parallel 0}=k_{\parallel 1}+k_{\parallel 2}$)
  for (a) dual-cascading and (b) inverse-cascading KAW triplets. 
  Black vector arrows represent the values of 
  frequency and parallel wavenumber for each modes $\bm{\Omega}_k=(\omega_{k},\bm{k})$
  and dashed lines constitute parallelograms for 
  different propagation direction of $\bm{\Omega}_2$.
  The modulus of slope of each dotted line is 
  proportional to $\alpha_k$, which increases monotonically with $k_{\perp}$.}
  \label{fig:Sketch_MatchCond}
\end{figure}

\section{Parametric Decay and Spectral Cascades} \label{sec:PDI}

Considering resonant three-KAW interaction process 
with frequency and wavenumber matching conditions 
$\bm{\Omega}_0=\bm{\Omega}_1+\bm{\Omega}_2$ satisfied, 
we look into the condition for spontaneous decay, with emphasis on the  energy flow 
among these three waves. 
Without loss of generality, taking $\bm{\Omega}_0$ as the pump wave, 
the frequency and parallel wavenumber matching conditions are illustrated in 
Fig. (\ref{fig:Sketch_MatchCond}), where cases for different propagation direction 
of $\bm{\Omega}_2$ with respect to $\bm{\Omega}_0$ are given.
It also implies the perpendicular wavenumber matching condition and 
embodies possible cascading behaviors in $k_{\perp}$, noting the dependence of the 
slope of $\omega_k/k_{\parallel}$ on $b_k$, which will be elaborated in  
the following analysis. 
Taking $\delta\phi_{k}(t)=\Phi_{k}(t)e^{-i\omega_{k}t}$ with $\Phi_{k}(t)$
being the slowly-varying amplitude, and noting 
$|\delta\phi_{k_0}|\gg |\delta\phi_{k_1}|,|\delta\phi_{k_2}|$,
we obtain the linearized coupled equations
\begin{eqnarray}
    b_{k_1}\epsilon_{Ak1}\Phi_{k_1}&=&-i(\Lambda_{k_2}^{k_1}/2\omega_{k_1})
    \beta_{k_2,k_0}\Phi_{k_2}^*\Phi_{k_0},\\
    b_{k_2}\epsilon_{Ak2}\Phi_{k_2}&=&-i(\Lambda_{k_2}^{k_1}/2\omega_{k_2})
    \beta_{k_0,k_1}\Phi_{k_0}\Phi_{k_1}^*, 
\end{eqnarray}
and $\Phi_{k_0}=\text{const}$. Expanding 
$\epsilon_{Ak}\simeq i\partial_{\omega_k}\epsilon_{Ak,r}(\partial_t-\gamma_{dk})$
with $\partial\epsilon_{Ak,r}/\partial\omega_k=2(1-\Gamma_k)/(b_k\omega_k)$, 
$\partial_t$ being the slowly temporal evolution and can be denoted as $\gamma$, 
and $\gamma_{dk}<0$ being the linear damping rate, we can rewrite
the coupled equations as 
\begin{eqnarray}
    (\partial_t-\gamma_{dk_1})\Phi_{k_1}&=&
    -\frac{\Lambda_{k_2}^{k_1}\beta_{k_2,k_0}}{4(1-\Gamma_{k_1})}\Phi_{k_2}^*\Phi_{k_0},\\
    (\partial_t-\gamma_{dk_2})\Phi_{k_2}&=&
    -\frac{\Lambda_{k_2}^{k_1}\beta_{k_0,k_1}}{4(1-\Gamma_{k_2})}\Phi_{k_0}\Phi_{k_1}^*.
\end{eqnarray}
The resultant parametric dispersion relation is given as
\begin{equation}\label{eqn:parametricD.R.}
    (\partial_t-\gamma_{dk_1})(\partial_t-\gamma_{dk_2})=
    \frac{(\Lambda_{k_2}^{k_1})^2\beta_{k_2,k_0}\beta_{k_0,k_1}}
    {16(1-\Gamma_{k_1})(1-\Gamma_{k_2})}|\Phi_{k_0}|^2,
\end{equation}
which then yields, the condition for  spontaneous decay with the $\gamma>0$ 
\begin{equation}\label{eqn:SpontDecay}
    \gamma_{ND}^2 \equiv \frac{(\Lambda_{k_2}^{k_1})^2
    \beta_{k_2,k_0}\beta_{k_0,k_1}|\Phi_{k_0}|^2}
    {16(1-\Gamma_{k_1})(1-\Gamma_{k_2})}> \gamma_{dk_1}\gamma_{dk_2}, 
\end{equation}
i.e., firstly, the sign of $\beta_{k_2,k_0}\beta_{k_0,k_1}$ to be positive, and 
secondly, the nonlinear drive intensity $\gamma^2_{ND}$ should exceed the threshold due to 
linear damping of sidebands, which determines the threshold on pump KAW amplitude. 
Note that, the threshold could also arise from the frequency mismatch, which is, 
however, not considered here. 

\begin{figure}[htbp!]
  \centering
  \subfigure{\label{fig:2a}\includegraphics[width=.99\linewidth]{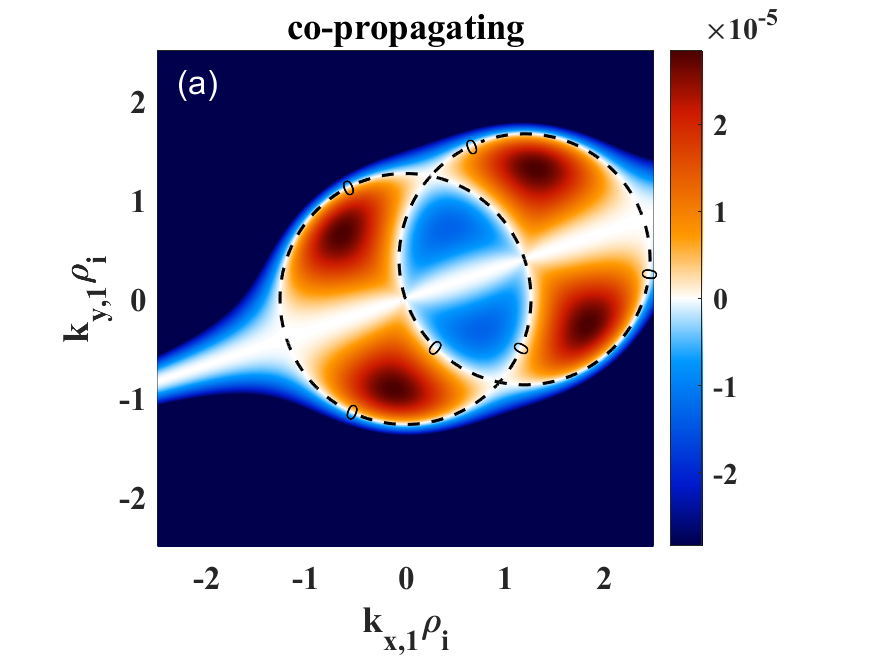}}  
  \subfigure{\label{fig:2b}\includegraphics[width=.99\linewidth]{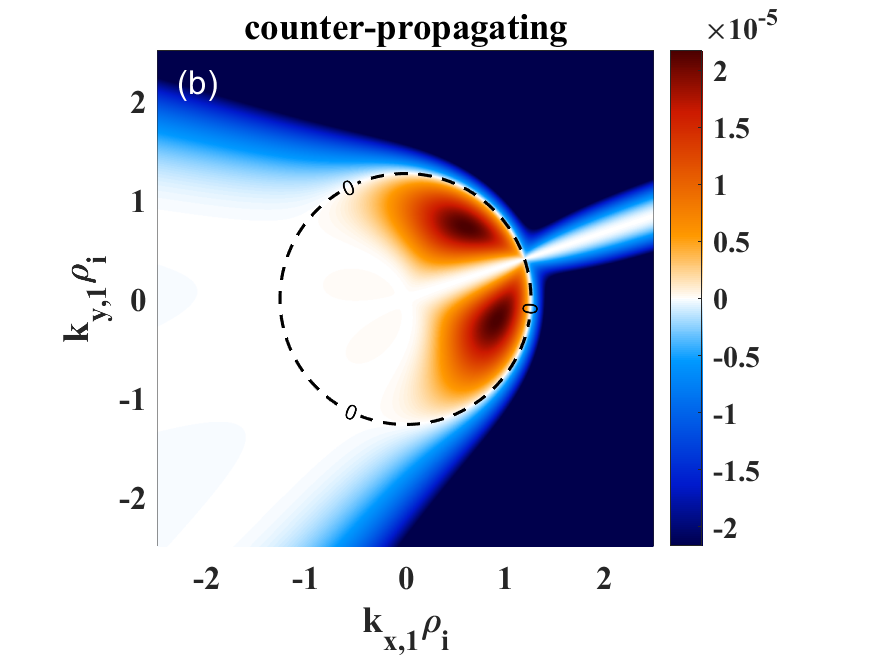}}
  \caption{Intensity profiles of $\gamma_{ND}^2/\Omega_i^2$
  for the sideband $\bm{\Omega}_2$ being (a) co-propagating 
  or (b) counter-propagating. Black dashed circles represent the 
  separatrix between positive and negative values.}
  \label{fig:NLDrive}
\end{figure}

The nonlinear drive $\gamma_{ND}^2/\Omega_i^2$ for co- and counter-propagating cases 
are visualized in two contour plots of Fig. \ref{fig:NLDrive}, with the horizontal 
and vertical axes being $k_{x1}\rho_i$ and $k_{y1}\rho_i$, respectively. 
The parameter used are 
$\beta_i=0.01$, $\tau =1$, pump KAW amplitude $|\delta B_{k_0}/B_0|=10^{-3}$, 
and $b_{k_0}=1.6$ with $k_{x,0}\rho_i=1.2$ and $k_{y,0}\rho_i=0.4$.
In Fig. \ref{fig:NLDrive}, red and blue colours correspond to positive and negative 
values, and the boundary between stable and unstable regions are indicated by 
black dashed circles. For all three KAWs being co-propagating, i.e.
a forward-scattering process as shown by the $(\Omega_0,\Omega_1,\Omega_2)$ 
combination in Fig. \ref{fig:1a}, the wavenumber spectrum exhibits 
a dual cascading character in $k_{\perp}$, similar to that of the DW described by 
Charney-Hasegawa-Mima equation,
which can be verified by the analytical expression 
$\beta_{k_2,k_0}\beta_{k_0,k_1}\propto (\alpha_{k_2}-\alpha_{k_0})
(\alpha_{k_0}-\alpha_{k_1})$, as well as the slopes of dotted lines 
in Fig. \ref{fig:1a}. While for the backward-scattering process with one sideband 
propagating in the direction opposite to that of the pump wave,  
as shown by the $(\Omega_0,\Omega_1',\Omega_2')$ combination in 
Fig. \ref{fig:1a} or Fig. \ref{fig:1b} 
(the difference is the slope of $\Omega_2$ with respect to the pump KAW $\Omega_0$),
the counter-propagating KAW triplet could be manifested as either dual-cascading 
or inverse-cascading within the circular region in Fig. \ref{fig:2b}, 
which is also confirmed by the analytical expression 
$\beta_{k_2,k_0}\beta_{k_0,k_1}\propto (\alpha_{k_0}-\alpha_{k_1})$. 
Above preliminary cascading properties obtained from spontaneous decay condition 
can also be identified in 
Fig. (\ref{fig:Sketch_MatchCond}), where the modulus of slope of 
each dotted line is proportional to $\alpha_k=\sqrt{\sigma_kb_k/(1-\Gamma_k)}$
and thus increases with $k_{\perp}$ monotonically.  
Furthermore, from the contour of $\gamma_{ND}^2/\Omega_i^2$, we conclude   
that the inverse cascade is dominant for counter-propagating case, 
since the interval with stronger nonlinear drive falls within the region 
of inverse cascading ($(\Omega_0,\Omega_1',\Omega_2')$ combination in 
Fig. \ref{fig:1b}). 
These different cascading behaviors in perpendicular wavenumber spectrum 
would further determine 
the stationary energy spectrum of KAW turbulence and 
have significant implication on cross-field transport. 

\begin{figure}[htbp!]
  \centering
  \includegraphics[width=.99\linewidth]{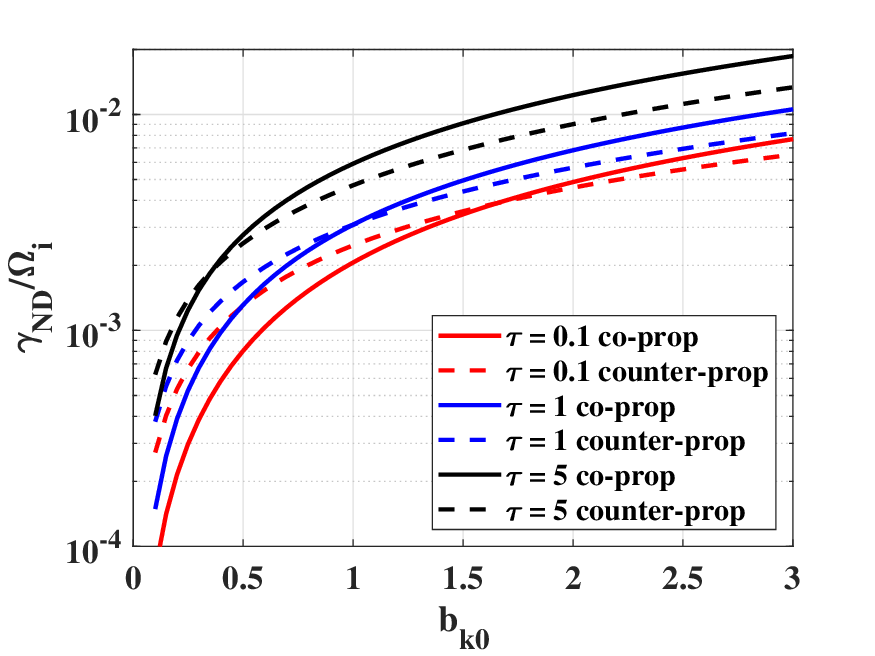}
  \caption{Parameter dependences of nonlinear drive $\gamma_{ND}/\Omega_i$ on 
  $b_{k_0}$ for different value of $\tau$.}
  \label{fig:NLDrive_b0&tau}
\end{figure}

For different value of temperature ratio $\tau$, 
we can evaluate the maximized nonlinear drive $\gamma_{ND}/\Omega_i$
for different $b_{k_0}$, as shown in Fig. \ref{fig:NLDrive_b0&tau}.
The overall tendency is that the maximized nonlinear drive increases with 
$b_{k_0}$ monotonically. For a specific $\tau$,  for relatively small $b_{k_0}$
decay is more effective into counter-streaming KAWs, i.e.   inverse cascade in $k_{\perp}$. While in the short-wavelength region with  relatively large $b_{k_0}$, 
the maximized nonlinear drive for co-propagating KAWs 
is slightly larger. Although the difference is not significant,
the consequences on resultant saturated spectrum could be worth exploring, 
considering different types of cascadings. 

\section{Conclusion and discussion} \label{sec:conclu}

In this work, the generalized nonlinear equation governing the resonant interaction 
among three kinetic Alfv\'en waves (KAWs) in the kinetic regime with 
$|k_{\perp}\rho_i|^2/|\omega/\Omega_i|\gtrsim\mathcal{O}(1)$ is derived 
using nonlinear gyrokinetic equation, motivated to obtain a generalized 
nonlinear theoretic framework that can be applied to strongly turbulent system, 
with the potential application to the spectral cascading of KAWs in nature and 
laboratory plasmas. A generalized expression of the nonlinear coupling coefficient 
with concrete physics meaning is derived, which can be conveniently used for the 
cases with the two beating KAWs being co- or counter-propagating along the 
equilibrium magnetic field. The present analysis assumed uniform plasma with 
$\beta\ll 1$ to simplify the analysis, however, extension to more realistic 
parameter regimes with plasma nonuniformity accounted for should be straightforward 
\cite{LChenPoP2022}. 

To reveal the crucial physics underlying the KAW spectral cascading, 
the condition for a pump KAW spontaneously decay into two sideband KAWs is analyzed. 
In the forward-scattering process with both sidebands being co-propagating 
with the pump KAW, it is found both analytically and numerically that, 
this corresponds to a dual cascading process in perpendicular wavenumber, 
similar to that of DW cascading described by the well-known Charney-Hasegawa-Mima 
equation. On the other hand, in the back-scattering process with one of the sidebands 
being counter-propagating with respect to the pump KAW, both dual- and inverse- cascading
are possible, and inverse-cascading may have a larger nonlinear cross-section. 
These aspects can be used in later analysis for the saturated KAW spectrum with 
linear drive/dissipation self-consistently accounted for. 

The general equation (\ref{eqn:NLEqn}) for resonant interactions among KAWs,  
with the nonlinear coupling coefficient given by Eq. (\ref{eqn:NLCoupCoeff}), 
is the most important result of the present work, and can be used to study 
the potentially strongly turbulent KAW nonlinear spectral cascading with 
$|\delta\omega/\omega|\sim\mathcal{O}(1)$, noting that $\omega\propto k_{\parallel}$ 
to the leading order for KAWs, with clear analogy to drift waves with 
$\omega\propto k_{\perp}$ that is imbedded in the spectral evolution described 
by Charney-Hasegawa-Mima equation. For further analytical progress, 
in the long wavelength limit with $b\ll1$, Eq. (\ref{eqn:NLEqn}) can be reduced to 
\begin{eqnarray}\label{eqn:kaw_co}
  \bm{\nabla}_{\perp}\cdot [\partial_t^2-v_A^2\rho_i^2\nabla_{\parallel}^2
  +(3/4+\tau)v_A^2\rho_i^4\nabla_{\parallel}^2\nabla_{\perp}^2]
  \bm{\nabla}_{\perp}\delta\phi \nonumber \\ 
  +(2c/B_0)\partial_t [(\hat{\bm{b}}\cdot\bm{\nabla}_{\perp}\delta\phi\times\bm{\nabla}_{\perp}) 
  \nabla_{\perp}^2\delta\phi] =0
\end{eqnarray}
and 
\begin{eqnarray}\label{eqn:kaw_counter}
  \bm{\nabla}_{\perp}\cdot [\partial_t^2-v_A^2\rho_i^2\nabla_{\parallel}^2
  +(3/4+\tau)v_A^2\rho_i^4\nabla_{\parallel}^2\nabla_{\perp}^2]
  \bm{\nabla}_{\perp}\delta\phi \nonumber \\ 
  +(3/4+\tau)(c\rho_i^2/B_0)\partial_t 
  [(\hat{\bm{b}}\cdot\bm{\nabla}_{\perp}\delta\phi\times\bm{\nabla}_{\perp}) 
  \nabla_{\perp}^4\delta\phi] =0
\end{eqnarray}
for counter- and co-propagating KAWs, respectively, and symmetrization has been 
applied in the derivation of these equations \cite{AHasegawaPoF1978}. 
They are capable of describing the self-consistent turbulent evolution 
of KAWs with multiple Fourier modes in long-wavelength regime. 
It is noteworthy that, the nonlinear term for counter-propagating case is similar 
to that of 2D Navier-Stokes/Euler equation and Charney-Hasegawa-Mima equation, 
while that for co-propagating case is quite different, suggesting both quantitatively 
and qualitatively different cascading dynamics. Therefore, the expected 
conservation laws, possible vortex solutions and direct numerical simulations of 
turbulent evolution will be given in future works. While applying the above 
Eqs. (\ref{eqn:kaw_co}) and (\ref{eqn:kaw_counter}), one needs to keep in mind that, 
they are derived assuming $b\ll 1$, which defines the parameter regime for 
their application.

As a final remark, while the interactions among KAWs are local in wavenumber space, 
it is expected that small-scale structures like convective cells \cite{FZoncaEPL2015} 
could be simultaneously excited, corresponding to zonal structures 
\cite{MRosenbluthPRL1998,FZoncaJPCS2021,PHDiamondPPCF2005}
prevalent in fusion and atmospheric plasmas, and these important physics 
are not included in the present model. Taking the zonal structure generation 
into account, one could study the nonlinear dynamics of KAWs, including the potential 
implications to fusion plasmas, could be analyzed in future projects. 

\begin{acknowledgments}
  We wish to thank Dr. Fulvio Zonca and Prof. Liu Chen for fruitful discussions. 
  Daily communications with Guangyu Wei and Ningfei Chen are also appreciated. 
  This work was supported by the National Science Foundation of China under 
  Grant Nos. 12275236 and 12261131622, and Italian Ministry for Foreign Affairs 
  and International Cooperation Project under Grant  No. CN23GR02.
\end{acknowledgments}

\section*{Author declarations}
\subsection*{Conflict of Interest}
The authors have no conflicts to disclose.

\section*{Data availability statement}
The data that support the findings of this study are 
available from the corresponding author upon reasonable request.

\bibliography{ref}

\begin{thebibliography}{51}%
\makeatletter
\providecommand \@ifxundefined [1]{%
 \@ifx{#1\undefined}
}%
\providecommand \@ifnum [1]{%
 \ifnum #1\expandafter \@firstoftwo
 \else \expandafter \@secondoftwo
 \fi
}%
\providecommand \@ifx [1]{%
 \ifx #1\expandafter \@firstoftwo
 \else \expandafter \@secondoftwo
 \fi
}%
\providecommand \natexlab [1]{#1}%
\providecommand \enquote  [1]{``#1''}%
\providecommand \bibnamefont  [1]{#1}%
\providecommand \bibfnamefont [1]{#1}%
\providecommand \citenamefont [1]{#1}%
\providecommand \href@noop [0]{\@secondoftwo}%
\providecommand \href [0]{\begingroup \@sanitize@url \@href}%
\providecommand \@href[1]{\@@startlink{#1}\@@href}%
\providecommand \@@href[1]{\endgroup#1\@@endlink}%
\providecommand \@sanitize@url [0]{\catcode `\\12\catcode `\$12\catcode `\&12\catcode `\#12\catcode `\^12\catcode `\_12\catcode `\%12\relax}%
\providecommand \@@startlink[1]{}%
\providecommand \@@endlink[0]{}%
\providecommand \url  [0]{\begingroup\@sanitize@url \@url }%
\providecommand \@url [1]{\endgroup\@href {#1}{\urlprefix }}%
\providecommand \urlprefix  [0]{URL }%
\providecommand \Eprint [0]{\href }%
\providecommand \doibase [0]{http://dx.doi.org/}%
\providecommand \selectlanguage [0]{\@gobble}%
\providecommand \bibinfo  [0]{\@secondoftwo}%
\providecommand \bibfield  [0]{\@secondoftwo}%
\providecommand \translation [1]{[#1]}%
\providecommand \BibitemOpen [0]{}%
\providecommand \bibitemStop [0]{}%
\providecommand \bibitemNoStop [0]{.\EOS\space}%
\providecommand \EOS [0]{\spacefactor3000\relax}%
\providecommand \BibitemShut  [1]{\csname bibitem#1\endcsname}%
\let\auto@bib@innerbib\@empty
\bibitem [{\citenamefont {Alfv{\'e}n}(1942)}]{HAlfvenNature1942}%
  \BibitemOpen
  \bibfield  {author} {\bibinfo {author} {\bibfnamefont {H.}~\bibnamefont {Alfv{\'e}n}},\ }\href@noop {} {\bibfield  {journal} {\bibinfo  {journal} {Nature}\ }\textbf {\bibinfo {volume} {150}},\ \bibinfo {pages} {405} (\bibinfo {year} {1942})}\BibitemShut {NoStop}%
\bibitem [{\citenamefont {Hasegawa}\ and\ \citenamefont {Chen}(1975)}]{AHasegawaPRL1975}%
  \BibitemOpen
  \bibfield  {author} {\bibinfo {author} {\bibfnamefont {A.}~\bibnamefont {Hasegawa}}\ and\ \bibinfo {author} {\bibfnamefont {L.}~\bibnamefont {Chen}},\ }\href@noop {} {\bibfield  {journal} {\bibinfo  {journal} {Physical Review Letters}\ }\textbf {\bibinfo {volume} {35}},\ \bibinfo {pages} {370} (\bibinfo {year} {1975})}\BibitemShut {NoStop}%
\bibitem [{\citenamefont {Grad}(1969)}]{HGradPT1969}%
  \BibitemOpen
  \bibfield  {author} {\bibinfo {author} {\bibfnamefont {H.}~\bibnamefont {Grad}},\ }\href {\doibase 10.1063/1.3035293} {\bibfield  {journal} {\bibinfo  {journal} {Physics Today}\ }\textbf {\bibinfo {volume} {22}},\ \bibinfo {pages} {34} (\bibinfo {year} {1969})}\BibitemShut {NoStop}%
\bibitem [{\citenamefont {Chen}\ and\ \citenamefont {Hasegawa}(1974{\natexlab{a}})}]{LChenPoF1974}%
  \BibitemOpen
  \bibfield  {author} {\bibinfo {author} {\bibfnamefont {L.}~\bibnamefont {Chen}}\ and\ \bibinfo {author} {\bibfnamefont {A.}~\bibnamefont {Hasegawa}},\ }\href@noop {} {\bibfield  {journal} {\bibinfo  {journal} {The Physics of Fluids}\ }\textbf {\bibinfo {volume} {17}},\ \bibinfo {pages} {1399} (\bibinfo {year} {1974}{\natexlab{a}})}\BibitemShut {NoStop}%
\bibitem [{\citenamefont {Hasegawa}\ and\ \citenamefont {Chen}(1974)}]{AHasegawaPRL1974}%
  \BibitemOpen
  \bibfield  {author} {\bibinfo {author} {\bibfnamefont {A.}~\bibnamefont {Hasegawa}}\ and\ \bibinfo {author} {\bibfnamefont {L.}~\bibnamefont {Chen}},\ }\href@noop {} {\bibfield  {journal} {\bibinfo  {journal} {Physical Review Letters}\ }\textbf {\bibinfo {volume} {32}},\ \bibinfo {pages} {454} (\bibinfo {year} {1974})}\BibitemShut {NoStop}%
\bibitem [{\citenamefont {Chen}\ and\ \citenamefont {Hasegawa}(1974{\natexlab{b}})}]{LChenJGR1974a}%
  \BibitemOpen
  \bibfield  {author} {\bibinfo {author} {\bibfnamefont {L.}~\bibnamefont {Chen}}\ and\ \bibinfo {author} {\bibfnamefont {A.}~\bibnamefont {Hasegawa}},\ }\href@noop {} {\bibfield  {journal} {\bibinfo  {journal} {Journal of Geophysical Research}\ }\textbf {\bibinfo {volume} {79}},\ \bibinfo {pages} {1024} (\bibinfo {year} {1974}{\natexlab{b}})}\BibitemShut {NoStop}%
\bibitem [{\citenamefont {Chen}\ and\ \citenamefont {Hasegawa}(1991)}]{LChenJGR1991}%
  \BibitemOpen
  \bibfield  {author} {\bibinfo {author} {\bibfnamefont {L.}~\bibnamefont {Chen}}\ and\ \bibinfo {author} {\bibfnamefont {A.}~\bibnamefont {Hasegawa}},\ }\href@noop {} {\bibfield  {journal} {\bibinfo  {journal} {Journal of Geophysical Research: Space Physics}\ }\textbf {\bibinfo {volume} {96}},\ \bibinfo {pages} {1503} (\bibinfo {year} {1991})}\BibitemShut {NoStop}%
\bibitem [{\citenamefont {Cranmer}\ and\ \citenamefont {Van~Ballegooijen}(2003)}]{SRCranmerApJ2003}%
  \BibitemOpen
  \bibfield  {author} {\bibinfo {author} {\bibfnamefont {S.}~\bibnamefont {Cranmer}}\ and\ \bibinfo {author} {\bibfnamefont {A.}~\bibnamefont {Van~Ballegooijen}},\ }\href@noop {} {\bibfield  {journal} {\bibinfo  {journal} {The Astrophysical Journal}\ }\textbf {\bibinfo {volume} {594}},\ \bibinfo {pages} {573} (\bibinfo {year} {2003})}\BibitemShut {NoStop}%
\bibitem [{\citenamefont {{Sagdeev}}\ and\ \citenamefont {{Galeev}}(1969)}]{RZSagdeev1969}%
  \BibitemOpen
  \bibfield  {author} {\bibinfo {author} {\bibfnamefont {R.~Z.}\ \bibnamefont {{Sagdeev}}}\ and\ \bibinfo {author} {\bibfnamefont {A.~A.}\ \bibnamefont {{Galeev}}},\ }\href@noop {} {\emph {\bibinfo {title} {Nonlinear plasma theory}}}\ (\bibinfo  {publisher} {AW Benjamin Inc.},\ \bibinfo {year} {1969})\BibitemShut {NoStop}%
\bibitem [{\citenamefont {Stenflo}(1994)}]{LStenfloPS1994}%
  \BibitemOpen
  \bibfield  {author} {\bibinfo {author} {\bibfnamefont {L.}~\bibnamefont {Stenflo}},\ }\href {\doibase 10.1088/0031-8949/1994/T50/002} {\bibfield  {journal} {\bibinfo  {journal} {Physica Scripta}\ }\textbf {\bibinfo {volume} {1994}},\ \bibinfo {pages} {15} (\bibinfo {year} {1994})}\BibitemShut {NoStop}%
\bibitem [{\citenamefont {Chen}\ and\ \citenamefont {Zonca}(2011)}]{LChenEPL2011}%
  \BibitemOpen
  \bibfield  {author} {\bibinfo {author} {\bibfnamefont {L.}~\bibnamefont {Chen}}\ and\ \bibinfo {author} {\bibfnamefont {F.}~\bibnamefont {Zonca}},\ }\href@noop {} {\bibfield  {journal} {\bibinfo  {journal} {Europhysics Letters}\ }\textbf {\bibinfo {volume} {96}},\ \bibinfo {pages} {35001} (\bibinfo {year} {2011})}\BibitemShut {NoStop}%
\bibitem [{\citenamefont {Hasegawa}\ and\ \citenamefont {Chen}(1976)}]{AHasegawaPoF1976}%
  \BibitemOpen
  \bibfield  {author} {\bibinfo {author} {\bibfnamefont {A.}~\bibnamefont {Hasegawa}}\ and\ \bibinfo {author} {\bibfnamefont {L.}~\bibnamefont {Chen}},\ }\href@noop {} {\bibfield  {journal} {\bibinfo  {journal} {The Physics of Fluids}\ }\textbf {\bibinfo {volume} {19}},\ \bibinfo {pages} {1924} (\bibinfo {year} {1976})}\BibitemShut {NoStop}%
\bibitem [{\citenamefont {Hasegawa}\ and\ \citenamefont {Mima}(1977)}]{AHasegawaPRL1977}%
  \BibitemOpen
  \bibfield  {author} {\bibinfo {author} {\bibfnamefont {A.}~\bibnamefont {Hasegawa}}\ and\ \bibinfo {author} {\bibfnamefont {K.}~\bibnamefont {Mima}},\ }\href@noop {} {\bibfield  {journal} {\bibinfo  {journal} {Physical Review Letters}\ }\textbf {\bibinfo {volume} {39}},\ \bibinfo {pages} {205} (\bibinfo {year} {1977})}\BibitemShut {NoStop}%
\bibitem [{\citenamefont {Hasegawa}\ and\ \citenamefont {Mima}(1978)}]{AHasegawaPoF1978}%
  \BibitemOpen
  \bibfield  {author} {\bibinfo {author} {\bibfnamefont {A.}~\bibnamefont {Hasegawa}}\ and\ \bibinfo {author} {\bibfnamefont {K.}~\bibnamefont {Mima}},\ }\href@noop {} {\bibfield  {journal} {\bibinfo  {journal} {The Physics of Fluids}\ }\textbf {\bibinfo {volume} {21}},\ \bibinfo {pages} {87} (\bibinfo {year} {1978})}\BibitemShut {NoStop}%
\bibitem [{\citenamefont {Rhines}(1975)}]{PBRhinesJFM1975}%
  \BibitemOpen
  \bibfield  {author} {\bibinfo {author} {\bibfnamefont {P.~B.}\ \bibnamefont {Rhines}},\ }\href {\doibase 10.1017/S0022112075001504} {\bibfield  {journal} {\bibinfo  {journal} {Journal of Fluid Mechanics}\ }\textbf {\bibinfo {volume} {69}},\ \bibinfo {pages} {417} (\bibinfo {year} {1975})}\BibitemShut {NoStop}%
\bibitem [{\citenamefont {Hasegawa}(1985)}]{AHasegawaAP1985}%
  \BibitemOpen
  \bibfield  {author} {\bibinfo {author} {\bibfnamefont {A.}~\bibnamefont {Hasegawa}},\ }\href {\doibase 10.1080/00018738500101721} {\bibfield  {journal} {\bibinfo  {journal} {Advances in Physics}\ }\textbf {\bibinfo {volume} {34}},\ \bibinfo {pages} {1} (\bibinfo {year} {1985})}\BibitemShut {NoStop}%
\bibitem [{\citenamefont {Boffetta}\ and\ \citenamefont {Ecke}(2012)}]{GBoffettaARFM2012}%
  \BibitemOpen
  \bibfield  {author} {\bibinfo {author} {\bibfnamefont {G.}~\bibnamefont {Boffetta}}\ and\ \bibinfo {author} {\bibfnamefont {R.~E.}\ \bibnamefont {Ecke}},\ }\href {\doibase 10.1146/annurev-fluid-120710-101240} {\bibfield  {journal} {\bibinfo  {journal} {Annual Review of Fluid Mechanics}\ }\textbf {\bibinfo {volume} {44}},\ \bibinfo {pages} {427} (\bibinfo {year} {2012})}\BibitemShut {NoStop}%
\bibitem [{\citenamefont {Kraichnan}\ and\ \citenamefont {Montgomery}(1980)}]{RHKraichnanRPP1980}%
  \BibitemOpen
  \bibfield  {author} {\bibinfo {author} {\bibfnamefont {R.~H.}\ \bibnamefont {Kraichnan}}\ and\ \bibinfo {author} {\bibfnamefont {D.}~\bibnamefont {Montgomery}},\ }\href {\doibase 10.1088/0034-4885/43/5/001} {\bibfield  {journal} {\bibinfo  {journal} {Reports on Progress in Physics}\ }\textbf {\bibinfo {volume} {43}},\ \bibinfo {pages} {547} (\bibinfo {year} {1980})}\BibitemShut {NoStop}%
\bibitem [{\citenamefont {Horbury}\ \emph {et~al.}(2005)\citenamefont {Horbury}, \citenamefont {Forman},\ and\ \citenamefont {Oughton}}]{TSHorburyPPCF2005}%
  \BibitemOpen
  \bibfield  {author} {\bibinfo {author} {\bibfnamefont {T.~S.}\ \bibnamefont {Horbury}}, \bibinfo {author} {\bibfnamefont {M.~A.}\ \bibnamefont {Forman}}, \ and\ \bibinfo {author} {\bibfnamefont {S.}~\bibnamefont {Oughton}},\ }\href {\doibase 10.1088/0741-3335/47/12B/S52} {\bibfield  {journal} {\bibinfo  {journal} {Plasma Physics and Controlled Fusion}\ }\textbf {\bibinfo {volume} {47}},\ \bibinfo {pages} {B703} (\bibinfo {year} {2005})}\BibitemShut {NoStop}%
\bibitem [{\citenamefont {Sahraoui}\ \emph {et~al.}(2009)\citenamefont {Sahraoui}, \citenamefont {Goldstein}, \citenamefont {Robert},\ and\ \citenamefont {Khotyaintsev}}]{FSahraouiPRL2009}%
  \BibitemOpen
  \bibfield  {author} {\bibinfo {author} {\bibfnamefont {F.}~\bibnamefont {Sahraoui}}, \bibinfo {author} {\bibfnamefont {M.~L.}\ \bibnamefont {Goldstein}}, \bibinfo {author} {\bibfnamefont {P.}~\bibnamefont {Robert}}, \ and\ \bibinfo {author} {\bibfnamefont {Y.~V.}\ \bibnamefont {Khotyaintsev}},\ }\href {\doibase 10.1103/PhysRevLett.102.231102} {\bibfield  {journal} {\bibinfo  {journal} {Physical Review Letters}\ }\textbf {\bibinfo {volume} {102}},\ \bibinfo {pages} {231102} (\bibinfo {year} {2009})}\BibitemShut {NoStop}%
\bibitem [{\citenamefont {Salem}\ \emph {et~al.}(2012)\citenamefont {Salem}, \citenamefont {Howes}, \citenamefont {Sundkvist}, \citenamefont {Bale}, \citenamefont {Chaston}, \citenamefont {Chen},\ and\ \citenamefont {Mozer}}]{CSSalemApJL2012}%
  \BibitemOpen
  \bibfield  {author} {\bibinfo {author} {\bibfnamefont {C.~S.}\ \bibnamefont {Salem}}, \bibinfo {author} {\bibfnamefont {G.~G.}\ \bibnamefont {Howes}}, \bibinfo {author} {\bibfnamefont {D.}~\bibnamefont {Sundkvist}}, \bibinfo {author} {\bibfnamefont {S.~D.}\ \bibnamefont {Bale}}, \bibinfo {author} {\bibfnamefont {C.~C.}\ \bibnamefont {Chaston}}, \bibinfo {author} {\bibfnamefont {C.~H.~K.}\ \bibnamefont {Chen}}, \ and\ \bibinfo {author} {\bibfnamefont {F.~S.}\ \bibnamefont {Mozer}},\ }\href {\doibase 10.1088/2041-8205/745/1/L9} {\bibfield  {journal} {\bibinfo  {journal} {The Astrophysical Journal Letters}\ }\textbf {\bibinfo {volume} {745}},\ \bibinfo {pages} {L9} (\bibinfo {year} {2012})}\BibitemShut {NoStop}%
\bibitem [{\citenamefont {Armstrong}\ \emph {et~al.}(1995)\citenamefont {Armstrong}, \citenamefont {Rickett},\ and\ \citenamefont {Spangler}}]{JWArmstrongApJ1995}%
  \BibitemOpen
  \bibfield  {author} {\bibinfo {author} {\bibfnamefont {J.}~\bibnamefont {Armstrong}}, \bibinfo {author} {\bibfnamefont {B.}~\bibnamefont {Rickett}}, \ and\ \bibinfo {author} {\bibfnamefont {S.}~\bibnamefont {Spangler}},\ }\href@noop {} {\bibfield  {journal} {\bibinfo  {journal} {The Astrophysical Journal}\ }\textbf {\bibinfo {volume} {443}},\ \bibinfo {pages} {209} (\bibinfo {year} {1995})}\BibitemShut {NoStop}%
\bibitem [{\citenamefont {Minter}\ and\ \citenamefont {Spangler}(1996)}]{AHMinterApJ1996}%
  \BibitemOpen
  \bibfield  {author} {\bibinfo {author} {\bibfnamefont {A.~H.}\ \bibnamefont {Minter}}\ and\ \bibinfo {author} {\bibfnamefont {S.~R.}\ \bibnamefont {Spangler}},\ }\href@noop {} {\bibfield  {journal} {\bibinfo  {journal} {The Astrophysical Journal}\ }\textbf {\bibinfo {volume} {458}},\ \bibinfo {pages} {194} (\bibinfo {year} {1996})}\BibitemShut {NoStop}%
\bibitem [{\citenamefont {Quataert}(1998)}]{EQuataertApJ1998}%
  \BibitemOpen
  \bibfield  {author} {\bibinfo {author} {\bibfnamefont {E.}~\bibnamefont {Quataert}},\ }\href@noop {} {\bibfield  {journal} {\bibinfo  {journal} {The Astrophysical Journal}\ }\textbf {\bibinfo {volume} {500}},\ \bibinfo {pages} {978} (\bibinfo {year} {1998})}\BibitemShut {NoStop}%
\bibitem [{\citenamefont {Quataert}\ and\ \citenamefont {Gruzinov}(1999)}]{EQuataertApJ1999}%
  \BibitemOpen
  \bibfield  {author} {\bibinfo {author} {\bibfnamefont {E.}~\bibnamefont {Quataert}}\ and\ \bibinfo {author} {\bibfnamefont {A.}~\bibnamefont {Gruzinov}},\ }\href@noop {} {\bibfield  {journal} {\bibinfo  {journal} {The Astrophysical Journal}\ }\textbf {\bibinfo {volume} {520}},\ \bibinfo {pages} {248} (\bibinfo {year} {1999})}\BibitemShut {NoStop}%
\bibitem [{\citenamefont {Frisch}(1995)}]{UFrischCUP1995}%
  \BibitemOpen
  \bibfield  {author} {\bibinfo {author} {\bibfnamefont {U.}~\bibnamefont {Frisch}},\ }\href@noop {} {\emph {\bibinfo {title} {Turbulence: the legacy of AN Kolmogorov}}}\ (\bibinfo  {publisher} {Cambridge university press},\ \bibinfo {year} {1995})\BibitemShut {NoStop}%
\bibitem [{\citenamefont {Krommes}(2012)}]{JAKrommesARFM2012}%
  \BibitemOpen
  \bibfield  {author} {\bibinfo {author} {\bibfnamefont {J.~A.}\ \bibnamefont {Krommes}},\ }\href@noop {} {\bibfield  {journal} {\bibinfo  {journal} {Annual Review of Fluid Mechanics}\ }\textbf {\bibinfo {volume} {44}},\ \bibinfo {pages} {175} (\bibinfo {year} {2012})}\BibitemShut {NoStop}%
\bibitem [{\citenamefont {Schekochihin}\ \emph {et~al.}(2008)\citenamefont {Schekochihin}, \citenamefont {Cowley}, \citenamefont {Dorland}, \citenamefont {Hammett}, \citenamefont {Howes}, \citenamefont {Plunk}, \citenamefont {Quataert},\ and\ \citenamefont {Tatsuno}}]{AASchekochihinPPCF2008}%
  \BibitemOpen
  \bibfield  {author} {\bibinfo {author} {\bibfnamefont {A.}~\bibnamefont {Schekochihin}}, \bibinfo {author} {\bibfnamefont {S.}~\bibnamefont {Cowley}}, \bibinfo {author} {\bibfnamefont {W.}~\bibnamefont {Dorland}}, \bibinfo {author} {\bibfnamefont {G.}~\bibnamefont {Hammett}}, \bibinfo {author} {\bibfnamefont {G.~G.}\ \bibnamefont {Howes}}, \bibinfo {author} {\bibfnamefont {G.}~\bibnamefont {Plunk}}, \bibinfo {author} {\bibfnamefont {E.}~\bibnamefont {Quataert}}, \ and\ \bibinfo {author} {\bibfnamefont {T.}~\bibnamefont {Tatsuno}},\ }\href@noop {} {\bibfield  {journal} {\bibinfo  {journal} {Plasma Physics and Controlled Fusion}\ }\textbf {\bibinfo {volume} {50}},\ \bibinfo {pages} {124024} (\bibinfo {year} {2008})}\BibitemShut {NoStop}%
\bibitem [{\citenamefont {Brizard}\ and\ \citenamefont {Hahm}(2007)}]{AJBrizardRMP2007}%
  \BibitemOpen
  \bibfield  {author} {\bibinfo {author} {\bibfnamefont {A.~J.}\ \bibnamefont {Brizard}}\ and\ \bibinfo {author} {\bibfnamefont {T.~S.}\ \bibnamefont {Hahm}},\ }\href@noop {} {\bibfield  {journal} {\bibinfo  {journal} {Reviews of Modern Physics}\ }\textbf {\bibinfo {volume} {79}},\ \bibinfo {pages} {421} (\bibinfo {year} {2007})}\BibitemShut {NoStop}%
\bibitem [{\citenamefont {Frieman}\ and\ \citenamefont {Chen}(1982)}]{EAFriemanPoF1982}%
  \BibitemOpen
  \bibfield  {author} {\bibinfo {author} {\bibfnamefont {E.}~\bibnamefont {Frieman}}\ and\ \bibinfo {author} {\bibfnamefont {L.}~\bibnamefont {Chen}},\ }\href@noop {} {\bibfield  {journal} {\bibinfo  {journal} {The Physics of Fluids}\ }\textbf {\bibinfo {volume} {25}},\ \bibinfo {pages} {502} (\bibinfo {year} {1982})}\BibitemShut {NoStop}%
\bibitem [{\citenamefont {Schekochihin}\ \emph {et~al.}(2009)\citenamefont {Schekochihin}, \citenamefont {Cowley}, \citenamefont {Dorland}, \citenamefont {Hammett}, \citenamefont {Howes}, \citenamefont {Quataert},\ and\ \citenamefont {Tatsuno}}]{AASchekochihinApJS2009}%
  \BibitemOpen
  \bibfield  {author} {\bibinfo {author} {\bibfnamefont {A.}~\bibnamefont {Schekochihin}}, \bibinfo {author} {\bibfnamefont {S.}~\bibnamefont {Cowley}}, \bibinfo {author} {\bibfnamefont {W.}~\bibnamefont {Dorland}}, \bibinfo {author} {\bibfnamefont {G.}~\bibnamefont {Hammett}}, \bibinfo {author} {\bibfnamefont {G.~G.}\ \bibnamefont {Howes}}, \bibinfo {author} {\bibfnamefont {E.}~\bibnamefont {Quataert}}, \ and\ \bibinfo {author} {\bibfnamefont {T.}~\bibnamefont {Tatsuno}},\ }\href@noop {} {\bibfield  {journal} {\bibinfo  {journal} {The Astrophysical Journal Supplement Series}\ }\textbf {\bibinfo {volume} {182}},\ \bibinfo {pages} {310} (\bibinfo {year} {2009})}\BibitemShut {NoStop}%
\bibitem [{\citenamefont {Howes}\ \emph {et~al.}(2006)\citenamefont {Howes}, \citenamefont {Cowley}, \citenamefont {Dorland}, \citenamefont {Hammett}, \citenamefont {Quataert},\ and\ \citenamefont {Schekochihin}}]{GGHowesApJ2006}%
  \BibitemOpen
  \bibfield  {author} {\bibinfo {author} {\bibfnamefont {G.~G.}\ \bibnamefont {Howes}}, \bibinfo {author} {\bibfnamefont {S.~C.}\ \bibnamefont {Cowley}}, \bibinfo {author} {\bibfnamefont {W.}~\bibnamefont {Dorland}}, \bibinfo {author} {\bibfnamefont {G.~W.}\ \bibnamefont {Hammett}}, \bibinfo {author} {\bibfnamefont {E.}~\bibnamefont {Quataert}}, \ and\ \bibinfo {author} {\bibfnamefont {A.~A.}\ \bibnamefont {Schekochihin}},\ }\href@noop {} {\bibfield  {journal} {\bibinfo  {journal} {The Astrophysical Journal}\ }\textbf {\bibinfo {volume} {651}},\ \bibinfo {pages} {590} (\bibinfo {year} {2006})}\BibitemShut {NoStop}%
\bibitem [{\citenamefont {Fasoli}\ \emph {et~al.}(2007)\citenamefont {Fasoli}, \citenamefont {Gormenzano}, \citenamefont {Berk}, \citenamefont {Breizman}, \citenamefont {Briguglio}, \citenamefont {Darrow}, \citenamefont {Gorelenkov}, \citenamefont {Heidbrink}, \citenamefont {Jaun}, \citenamefont {Konovalov} \emph {et~al.}}]{AFasoliNF2007}%
  \BibitemOpen
  \bibfield  {author} {\bibinfo {author} {\bibfnamefont {A.}~\bibnamefont {Fasoli}}, \bibinfo {author} {\bibfnamefont {C.}~\bibnamefont {Gormenzano}}, \bibinfo {author} {\bibfnamefont {H.}~\bibnamefont {Berk}}, \bibinfo {author} {\bibfnamefont {B.}~\bibnamefont {Breizman}}, \bibinfo {author} {\bibfnamefont {S.}~\bibnamefont {Briguglio}}, \bibinfo {author} {\bibfnamefont {D.}~\bibnamefont {Darrow}}, \bibinfo {author} {\bibfnamefont {N.}~\bibnamefont {Gorelenkov}}, \bibinfo {author} {\bibfnamefont {W.}~\bibnamefont {Heidbrink}}, \bibinfo {author} {\bibfnamefont {A.}~\bibnamefont {Jaun}}, \bibinfo {author} {\bibfnamefont {S.}~\bibnamefont {Konovalov}},  \emph {et~al.},\ }\href@noop {} {\bibfield  {journal} {\bibinfo  {journal} {Nuclear Fusion}\ }\textbf {\bibinfo {volume} {47}},\ \bibinfo {pages} {S264} (\bibinfo {year} {2007})}\BibitemShut {NoStop}%
\bibitem [{\citenamefont {Chen}\ and\ \citenamefont {Zonca}(2016)}]{LChenRMP2016}%
  \BibitemOpen
  \bibfield  {author} {\bibinfo {author} {\bibfnamefont {L.}~\bibnamefont {Chen}}\ and\ \bibinfo {author} {\bibfnamefont {F.}~\bibnamefont {Zonca}},\ }\href@noop {} {\bibfield  {journal} {\bibinfo  {journal} {Reviews of Modern Physics}\ }\textbf {\bibinfo {volume} {88}},\ \bibinfo {pages} {015008} (\bibinfo {year} {2016})}\BibitemShut {NoStop}%
\bibitem [{\citenamefont {Liu}\ \emph {et~al.}(2022)\citenamefont {Liu}, \citenamefont {Wei}, \citenamefont {Lin}, \citenamefont {Brochard}, \citenamefont {Choi}, \citenamefont {Heidbrink}, \citenamefont {Nicolau},\ and\ \citenamefont {McKee}}]{PLiuPRL2022}%
  \BibitemOpen
  \bibfield  {author} {\bibinfo {author} {\bibfnamefont {P.}~\bibnamefont {Liu}}, \bibinfo {author} {\bibfnamefont {X.}~\bibnamefont {Wei}}, \bibinfo {author} {\bibfnamefont {Z.}~\bibnamefont {Lin}}, \bibinfo {author} {\bibfnamefont {G.}~\bibnamefont {Brochard}}, \bibinfo {author} {\bibfnamefont {G.~J.}\ \bibnamefont {Choi}}, \bibinfo {author} {\bibfnamefont {W.~W.}\ \bibnamefont {Heidbrink}}, \bibinfo {author} {\bibfnamefont {J.~H.}\ \bibnamefont {Nicolau}}, \ and\ \bibinfo {author} {\bibfnamefont {G.~R.}\ \bibnamefont {McKee}},\ }\href {\doibase 10.1103/PhysRevLett.128.185001} {\bibfield  {journal} {\bibinfo  {journal} {Physical Review Letters}\ }\textbf {\bibinfo {volume} {128}},\ \bibinfo {pages} {185001} (\bibinfo {year} {2022})}\BibitemShut {NoStop}%
\bibitem [{\citenamefont {Liu}\ \emph {et~al.}(2023)\citenamefont {Liu}, \citenamefont {Wei}, \citenamefont {Lin}, \citenamefont {Brochard}, \citenamefont {Choi},\ and\ \citenamefont {Nicolau}}]{PLiuRMPP2023}%
  \BibitemOpen
  \bibfield  {author} {\bibinfo {author} {\bibfnamefont {P.}~\bibnamefont {Liu}}, \bibinfo {author} {\bibfnamefont {X.}~\bibnamefont {Wei}}, \bibinfo {author} {\bibfnamefont {Z.}~\bibnamefont {Lin}}, \bibinfo {author} {\bibfnamefont {G.}~\bibnamefont {Brochard}}, \bibinfo {author} {\bibfnamefont {G.}~\bibnamefont {Choi}}, \ and\ \bibinfo {author} {\bibfnamefont {J.}~\bibnamefont {Nicolau}},\ }\href@noop {} {\bibfield  {journal} {\bibinfo  {journal} {Reviews of Modern Plasma Physics}\ }\textbf {\bibinfo {volume} {7}},\ \bibinfo {pages} {15} (\bibinfo {year} {2023})}\BibitemShut {NoStop}%
\bibitem [{\citenamefont {Ye}\ \emph {et~al.}(2022)\citenamefont {Ye}, \citenamefont {Chen},\ and\ \citenamefont {Fu}}]{LYeNF2023}%
  \BibitemOpen
  \bibfield  {author} {\bibinfo {author} {\bibfnamefont {L.}~\bibnamefont {Ye}}, \bibinfo {author} {\bibfnamefont {Y.}~\bibnamefont {Chen}}, \ and\ \bibinfo {author} {\bibfnamefont {G.}~\bibnamefont {Fu}},\ }\href {\doibase 10.1088/1741-4326/aca9df} {\bibfield  {journal} {\bibinfo  {journal} {Nuclear Fusion}\ }\textbf {\bibinfo {volume} {63}},\ \bibinfo {pages} {026004} (\bibinfo {year} {2022})}\BibitemShut {NoStop}%
\bibitem [{\citenamefont {Wei}\ \emph {et~al.}(2022)\citenamefont {Wei}, \citenamefont {Wang}, \citenamefont {Chen}, \citenamefont {Zonca},\ and\ \citenamefont {Qiu}}]{SWeiNF2022}%
  \BibitemOpen
  \bibfield  {author} {\bibinfo {author} {\bibfnamefont {S.}~\bibnamefont {Wei}}, \bibinfo {author} {\bibfnamefont {T.}~\bibnamefont {Wang}}, \bibinfo {author} {\bibfnamefont {L.}~\bibnamefont {Chen}}, \bibinfo {author} {\bibfnamefont {F.}~\bibnamefont {Zonca}}, \ and\ \bibinfo {author} {\bibfnamefont {Z.}~\bibnamefont {Qiu}},\ }\href@noop {} {\bibfield  {journal} {\bibinfo  {journal} {Nuclear Fusion}\ }\textbf {\bibinfo {volume} {62}},\ \bibinfo {pages} {126038} (\bibinfo {year} {2022})}\BibitemShut {NoStop}%
\bibitem [{\citenamefont {Todo}(2018)}]{YTodoRMPP2018}%
  \BibitemOpen
  \bibfield  {author} {\bibinfo {author} {\bibfnamefont {Y.}~\bibnamefont {Todo}},\ }\href@noop {} {\bibfield  {journal} {\bibinfo  {journal} {Reviews of Modern Plasma Physics}\ }\textbf {\bibinfo {volume} {3}},\ \bibinfo {pages} {1} (\bibinfo {year} {2018})}\BibitemShut {NoStop}%
\bibitem [{\citenamefont {Chen}\ and\ \citenamefont {Zonca}(2013)}]{LChenPoP2013}%
  \BibitemOpen
  \bibfield  {author} {\bibinfo {author} {\bibfnamefont {L.}~\bibnamefont {Chen}}\ and\ \bibinfo {author} {\bibfnamefont {F.}~\bibnamefont {Zonca}},\ }\href@noop {} {\bibfield  {journal} {\bibinfo  {journal} {Physics of Plasmas}\ }\textbf {\bibinfo {volume} {20}} (\bibinfo {year} {2013})}\BibitemShut {NoStop}%
\bibitem [{\citenamefont {Chen}\ \emph {et~al.}(2001)\citenamefont {Chen}, \citenamefont {Lin}, \citenamefont {White},\ and\ \citenamefont {Zonca}}]{LChenNF2001}%
  \BibitemOpen
  \bibfield  {author} {\bibinfo {author} {\bibfnamefont {L.}~\bibnamefont {Chen}}, \bibinfo {author} {\bibfnamefont {Z.}~\bibnamefont {Lin}}, \bibinfo {author} {\bibfnamefont {R.~B.}\ \bibnamefont {White}}, \ and\ \bibinfo {author} {\bibfnamefont {F.}~\bibnamefont {Zonca}},\ }\href@noop {} {\bibfield  {journal} {\bibinfo  {journal} {Nuclear Fusion}\ }\textbf {\bibinfo {volume} {41}},\ \bibinfo {pages} {747} (\bibinfo {year} {2001})}\BibitemShut {NoStop}%
\bibitem [{\citenamefont {Hahm}\ and\ \citenamefont {Chen}(1995)}]{TSHahmPRL1995}%
  \BibitemOpen
  \bibfield  {author} {\bibinfo {author} {\bibfnamefont {T.}~\bibnamefont {Hahm}}\ and\ \bibinfo {author} {\bibfnamefont {L.}~\bibnamefont {Chen}},\ }\href@noop {} {\bibfield  {journal} {\bibinfo  {journal} {Physical Review Letters}\ }\textbf {\bibinfo {volume} {74}},\ \bibinfo {pages} {266} (\bibinfo {year} {1995})}\BibitemShut {NoStop}%
\bibitem [{\citenamefont {Bale}\ \emph {et~al.}(2005)\citenamefont {Bale}, \citenamefont {Kellogg}, \citenamefont {Mozer}, \citenamefont {Horbury},\ and\ \citenamefont {Reme}}]{SDBalePRL2005}%
  \BibitemOpen
  \bibfield  {author} {\bibinfo {author} {\bibfnamefont {S.}~\bibnamefont {Bale}}, \bibinfo {author} {\bibfnamefont {P.}~\bibnamefont {Kellogg}}, \bibinfo {author} {\bibfnamefont {F.}~\bibnamefont {Mozer}}, \bibinfo {author} {\bibfnamefont {T.}~\bibnamefont {Horbury}}, \ and\ \bibinfo {author} {\bibfnamefont {H.}~\bibnamefont {Reme}},\ }\href@noop {} {\bibfield  {journal} {\bibinfo  {journal} {Physical Review Letters}\ }\textbf {\bibinfo {volume} {94}},\ \bibinfo {pages} {215002} (\bibinfo {year} {2005})}\BibitemShut {NoStop}%
\bibitem [{\citenamefont {Howes}\ \emph {et~al.}(2008)\citenamefont {Howes}, \citenamefont {Dorland}, \citenamefont {Cowley}, \citenamefont {Hammett}, \citenamefont {Quataert}, \citenamefont {Schekochihin},\ and\ \citenamefont {Tatsuno}}]{GGHowesPRL2008}%
  \BibitemOpen
  \bibfield  {author} {\bibinfo {author} {\bibfnamefont {G.}~\bibnamefont {Howes}}, \bibinfo {author} {\bibfnamefont {W.}~\bibnamefont {Dorland}}, \bibinfo {author} {\bibfnamefont {S.}~\bibnamefont {Cowley}}, \bibinfo {author} {\bibfnamefont {G.}~\bibnamefont {Hammett}}, \bibinfo {author} {\bibfnamefont {E.}~\bibnamefont {Quataert}}, \bibinfo {author} {\bibfnamefont {A.}~\bibnamefont {Schekochihin}}, \ and\ \bibinfo {author} {\bibfnamefont {T.}~\bibnamefont {Tatsuno}},\ }\href@noop {} {\bibfield  {journal} {\bibinfo  {journal} {Physical Review Letters}\ }\textbf {\bibinfo {volume} {100}},\ \bibinfo {pages} {065004} (\bibinfo {year} {2008})}\BibitemShut {NoStop}%
\bibitem [{\citenamefont {Howes}\ \emph {et~al.}(2011)\citenamefont {Howes}, \citenamefont {TenBarge}, \citenamefont {Dorland}, \citenamefont {Quataert}, \citenamefont {Schekochihin}, \citenamefont {Numata},\ and\ \citenamefont {Tatsuno}}]{GGHowesPRL2011}%
  \BibitemOpen
  \bibfield  {author} {\bibinfo {author} {\bibfnamefont {G.~G.}\ \bibnamefont {Howes}}, \bibinfo {author} {\bibfnamefont {J.~M.}\ \bibnamefont {TenBarge}}, \bibinfo {author} {\bibfnamefont {W.}~\bibnamefont {Dorland}}, \bibinfo {author} {\bibfnamefont {E.}~\bibnamefont {Quataert}}, \bibinfo {author} {\bibfnamefont {A.~A.}\ \bibnamefont {Schekochihin}}, \bibinfo {author} {\bibfnamefont {R.}~\bibnamefont {Numata}}, \ and\ \bibinfo {author} {\bibfnamefont {T.}~\bibnamefont {Tatsuno}},\ }\href@noop {} {\bibfield  {journal} {\bibinfo  {journal} {Physical Review Letters}\ }\textbf {\bibinfo {volume} {107}},\ \bibinfo {pages} {035004} (\bibinfo {year} {2011})}\BibitemShut {NoStop}%
\bibitem [{\citenamefont {Voitenko}(1998)}]{YMVoitenkoJPP1998}%
  \BibitemOpen
  \bibfield  {author} {\bibinfo {author} {\bibfnamefont {Y.~M.}\ \bibnamefont {Voitenko}},\ }\href@noop {} {\bibfield  {journal} {\bibinfo  {journal} {Journal of Plasma Physics}\ }\textbf {\bibinfo {volume} {60}},\ \bibinfo {pages} {497} (\bibinfo {year} {1998})}\BibitemShut {NoStop}%
\bibitem [{\citenamefont {Chen}\ \emph {et~al.}(2022)\citenamefont {Chen}, \citenamefont {Qiu},\ and\ \citenamefont {Zonca}}]{LChenPoP2022}%
  \BibitemOpen
  \bibfield  {author} {\bibinfo {author} {\bibfnamefont {L.}~\bibnamefont {Chen}}, \bibinfo {author} {\bibfnamefont {Z.}~\bibnamefont {Qiu}}, \ and\ \bibinfo {author} {\bibfnamefont {F.}~\bibnamefont {Zonca}},\ }\href@noop {} {\bibfield  {journal} {\bibinfo  {journal} {Physics of Plasmas}\ }\textbf {\bibinfo {volume} {29}} (\bibinfo {year} {2022})}\BibitemShut {NoStop}%
\bibitem [{\citenamefont {Zonca}\ \emph {et~al.}(2015)\citenamefont {Zonca}, \citenamefont {Lin},\ and\ \citenamefont {Chen}}]{FZoncaEPL2015}%
  \BibitemOpen
  \bibfield  {author} {\bibinfo {author} {\bibfnamefont {F.}~\bibnamefont {Zonca}}, \bibinfo {author} {\bibfnamefont {Y.}~\bibnamefont {Lin}}, \ and\ \bibinfo {author} {\bibfnamefont {L.}~\bibnamefont {Chen}},\ }\href@noop {} {\bibfield  {journal} {\bibinfo  {journal} {Europhysics Letters}\ }\textbf {\bibinfo {volume} {112}},\ \bibinfo {pages} {65001} (\bibinfo {year} {2015})}\BibitemShut {NoStop}%
\bibitem [{\citenamefont {Rosenbluth}\ and\ \citenamefont {Hinton}(1998)}]{MRosenbluthPRL1998}%
  \BibitemOpen
  \bibfield  {author} {\bibinfo {author} {\bibfnamefont {M.~N.}\ \bibnamefont {Rosenbluth}}\ and\ \bibinfo {author} {\bibfnamefont {F.~L.}\ \bibnamefont {Hinton}},\ }\href@noop {} {\bibfield  {journal} {\bibinfo  {journal} {Physical Review Letters}\ }\textbf {\bibinfo {volume} {80}},\ \bibinfo {pages} {724} (\bibinfo {year} {1998})}\BibitemShut {NoStop}%
\bibitem [{\citenamefont {Zonca}\ \emph {et~al.}(2021)\citenamefont {Zonca}, \citenamefont {Chen}, \citenamefont {Falessi},\ and\ \citenamefont {Qiu}}]{FZoncaJPCS2021}%
  \BibitemOpen
  \bibfield  {author} {\bibinfo {author} {\bibfnamefont {F.}~\bibnamefont {Zonca}}, \bibinfo {author} {\bibfnamefont {L.}~\bibnamefont {Chen}}, \bibinfo {author} {\bibfnamefont {M.}~\bibnamefont {Falessi}}, \ and\ \bibinfo {author} {\bibfnamefont {Z.}~\bibnamefont {Qiu}},\ }\href@noop {} {\bibfield  {journal} {\bibinfo  {journal} {Journal of Physics: Conference Series}\ }\textbf {\bibinfo {volume} {1785}},\ \bibinfo {pages} {012005} (\bibinfo {year} {2021})}\BibitemShut {NoStop}%
\bibitem [{\citenamefont {Diamond}\ \emph {et~al.}(2005)\citenamefont {Diamond}, \citenamefont {Itoh}, \citenamefont {Itoh},\ and\ \citenamefont {Hahm}}]{PHDiamondPPCF2005}%
  \BibitemOpen
  \bibfield  {author} {\bibinfo {author} {\bibfnamefont {P.~H.}\ \bibnamefont {Diamond}}, \bibinfo {author} {\bibfnamefont {S.-I.}\ \bibnamefont {Itoh}}, \bibinfo {author} {\bibfnamefont {K.}~\bibnamefont {Itoh}}, \ and\ \bibinfo {author} {\bibfnamefont {T.~S.}\ \bibnamefont {Hahm}},\ }\href {\doibase 10.1088/0741-3335/47/5/R01} {\bibfield  {journal} {\bibinfo  {journal} {Plasma Physics and Controlled Fusion}\ }\textbf {\bibinfo {volume} {47}},\ \bibinfo {pages} {R35} (\bibinfo {year} {2005})}\BibitemShut {NoStop}%
\end{thebibliography}%

\end{document}